\documentclass[11pt,a4paper]{article}
\usepackage{graphicx,amsfonts}

\newcommand{\rv}{{\mathbf r}}
\newcommand{\kv}{{\mathbf k}}
\newcommand{\bv}{{\mathbf b}}
\newcommand{\Zset}{\mathbb{Z}}

\begin{document}


\begin{title}
{Dislocation Dynamics in a Dodecagonal Quasiperiodic Structure}
\end{title}
\author{Gilad Barak and Ron Lifshitz \\\\ School of Physics and
Astronomy\\  Raymond and Beverly Sackler Faculty of Exact
Sciences\\ Tel Aviv University, Tel Aviv 69978, Israel}
\date{May 15, 2005}
\maketitle

\begin{abstract}
  We have developed a set of numerical tools for the quantitative
  analysis of defect dynamics in quasiperiodic structures. We have
  applied these tools to study dislocation motion in the dynamical
  equation of Lifshitz and Petrich [Phys.\ Rev.\ Lett.\ {\bf 79}
  (1997) 1261] whose steady state solutions include a quasiperiodic
  structure with dodecagonal symmetry. Arbitrary dislocations,
  parameterized by the homotopy group of the D-torus, are injected as
  initial conditions and quantitatively followed as the equation
  evolves in real time. We show that for strong diffusion the results
  for dislocation climb velocity are similar for the dodecagonal and
  the hexagonal patterns, but that for weak diffusion the
  dodecagonal pattern exhibits a unique pinning of the dislocation,
  reflecting its quasiperiodic nature.
 \end{abstract}

\section{Introduction}

It was realized long ago that dislocations play an important role in
determining the mechanical properties of a crystal---whether periodic
or not. As a consequence, ever since the discovery of quasicrystals
much effort has been invested in the experimental and the theoretical
study of their dislocations~\cite{urban,trebin,woll}, leading to
interesting observations such as the special role played by phasons in
the motion of dislocations. 

Here we study the behavior of dislocations in a model system based on
the dynamical equation of Lifshitz and Petrich~\cite[henceforth
LP]{faraday}, which is a modification of the well-known
Swift-Hohenberg equation~\cite{swift}. The dynamics of the LP equation
is relaxational, $\partial_t \rho = -\delta {\cal F} / \delta \rho$,
driving a continuous 2-dimensional density field $\rho(x,y,t)$ towards
a minimum of the Lyapunov functional, or ``effective free energy'',
\begin{equation}
 \label{eq:lyapunov}
 {\cal F}\{\rho\} = \int \!dx\, dy\, \bigl\{- \frac{\varepsilon}{2}
 \rho^2 + \frac{c}{2} 
 [(\nabla^2+1)(\nabla^2+q^2)\rho]^2 
 - \frac13 \rho^3 + \frac14 \rho^4 \bigr\},
\end{equation}
yielding a dynamical equation of the form
\begin{equation}
 \label{lpeqn}
 \partial_t \rho = \varepsilon \rho - c(\nabla^2 + 1)^2(\nabla^2 +q^2)^2 \rho
 +\rho^2 - \rho^3.
\end{equation}
Among their results, LP showed that if $q=2\cos(\pi/4)$,
$q=2\cos(\pi/6)$, or $q=2\cos(\pi/12)$ and $\varepsilon$ is
sufficiently small, stable steady-state patterns are obtained (from
random initial conditions) with long-range order and square,
hexagonal, or dodecagonal symmetry, respectively.  Although the LP
equation cannot describe the dynamics of real solid-state
quasicrystals and is probably more appropriate for soft matter or
fluid phenomena such as Faraday waves, it offers a unique opportunity
for the quantitative study of dislocations in a system---exhibiting
both periodic and quasiperiodic long-range order---whose dynamics is
exactly known.

\section{Injection and tracking of dislocations}

The Lyapunov functional $\cal F$ in~(\ref{eq:lyapunov}) is clearly
invariant under any translation or rotation of space. The steady-state
solutions that are obtained are symmetry-broken ground states of $\cal
F$, whose Fourier transform has the form
\begin{equation}
  \label{eq:fourier}
  \rho(\rv) = \sum_{\kv\in L} \rho(\kv) e^{i\kv\cdot\rv},
\end{equation}
where the (reciprocal) lattice $L$ is a finitely generated
$\Zset$-module, {\it i.e.}  it can be expressed as the set of all
integral linear combination of a finite number $D$ of $d$-dimensional
wave vectors. In the special case where $D$, called the rank of the
structure, is equal to the physical dimension $d$ (here $d=2$), the
structure is periodic.

Any particular ground state $\rho({\rv})$ of $\cal F$ is {\it
  indistinguishable\/} from a whole set of ground states that are
related by so-called gauge transformations
\begin{equation}
  \label{eq:indist2}
  \forall \kv\in L:\quad \rho'(\kv) = e^{2\pi i\chi(\kv)}\rho(\kv),
\end{equation}
where $\chi(\kv)$, called a {\it gauge-function,} has the property
that $\chi(\kv_1+\kv_2) = \chi(\kv_1) + \chi(\kv_2)$, possibly to
within an additive integer, whenever $\kv_1$ and $\kv_2$ are in $L$.
As described in detail by Dr\"ager and Mermin~\cite{jorg}, gauge
functions form a vector space $V^*$ of all real-valued linear
functions on the lattice $L$. Because $L$ has rank $D$, any linear
function is completely specified by giving its values on $D$
integrally-independent lattice vectors. The space $V^*$ is therefore a
$D$-dimensional vector space over the real numbers. The space $V^*$
contains, as a subset, all the integral-valued linear functions on
$L$, denoted by $L^*$. Gauge functions in $L^*$ leave the ground-state
density invariant. Gauge functions that belong to the quotient space
$V^*/L^*$ take the ground state described by $\rho$ into a different,
yet indistinguishable, ground state described by some other density
function $\rho'$. Thus, one can parameterize all the related ground
states of $\cal F$ on a simple $D$-torus---the order parameter space
$V^*/L^*$.

\begin{figure}[tb]
\begin{center}
\includegraphics[height=7cm]{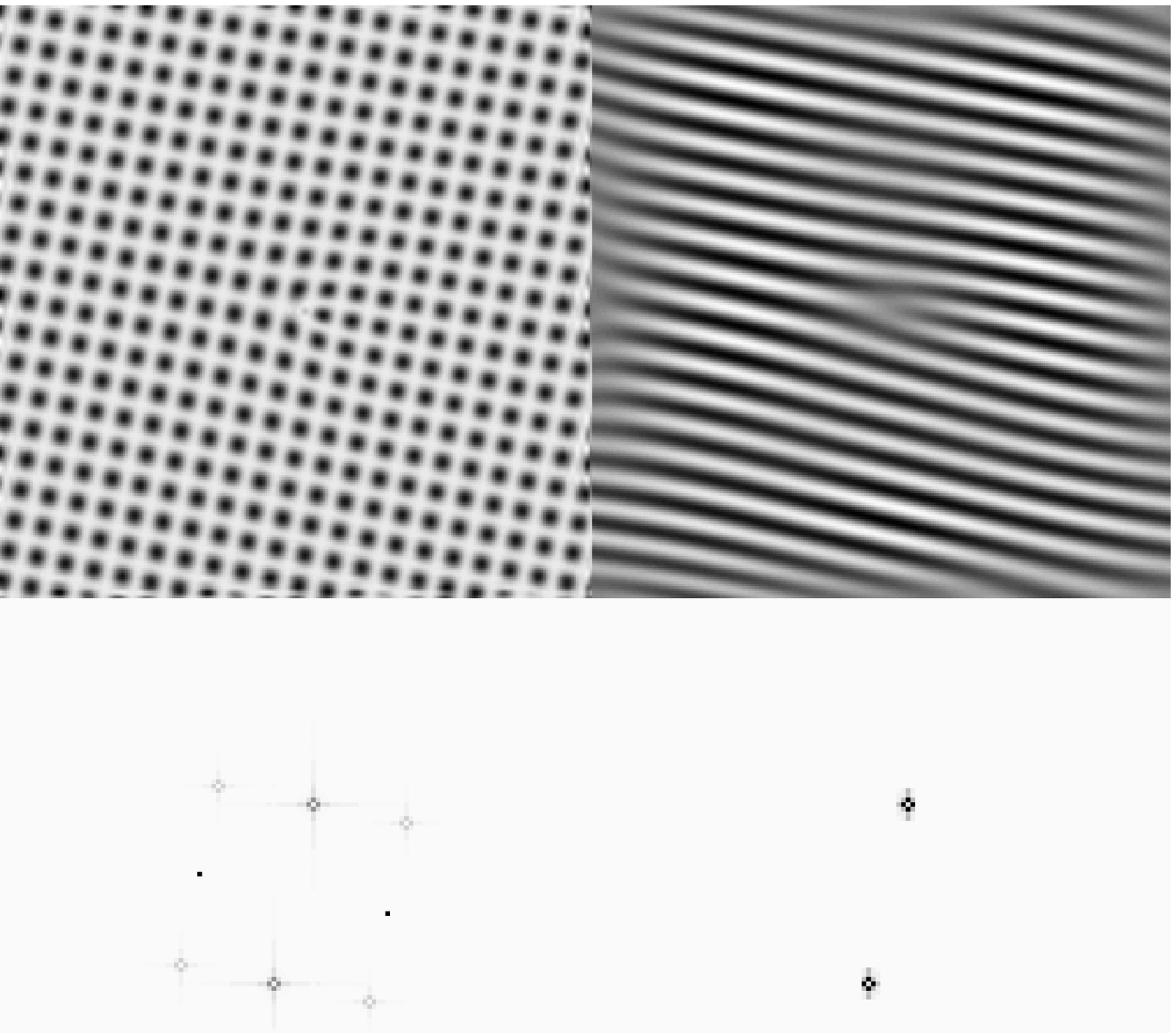}
\hspace{0.5cm}
\includegraphics[height=7cm]{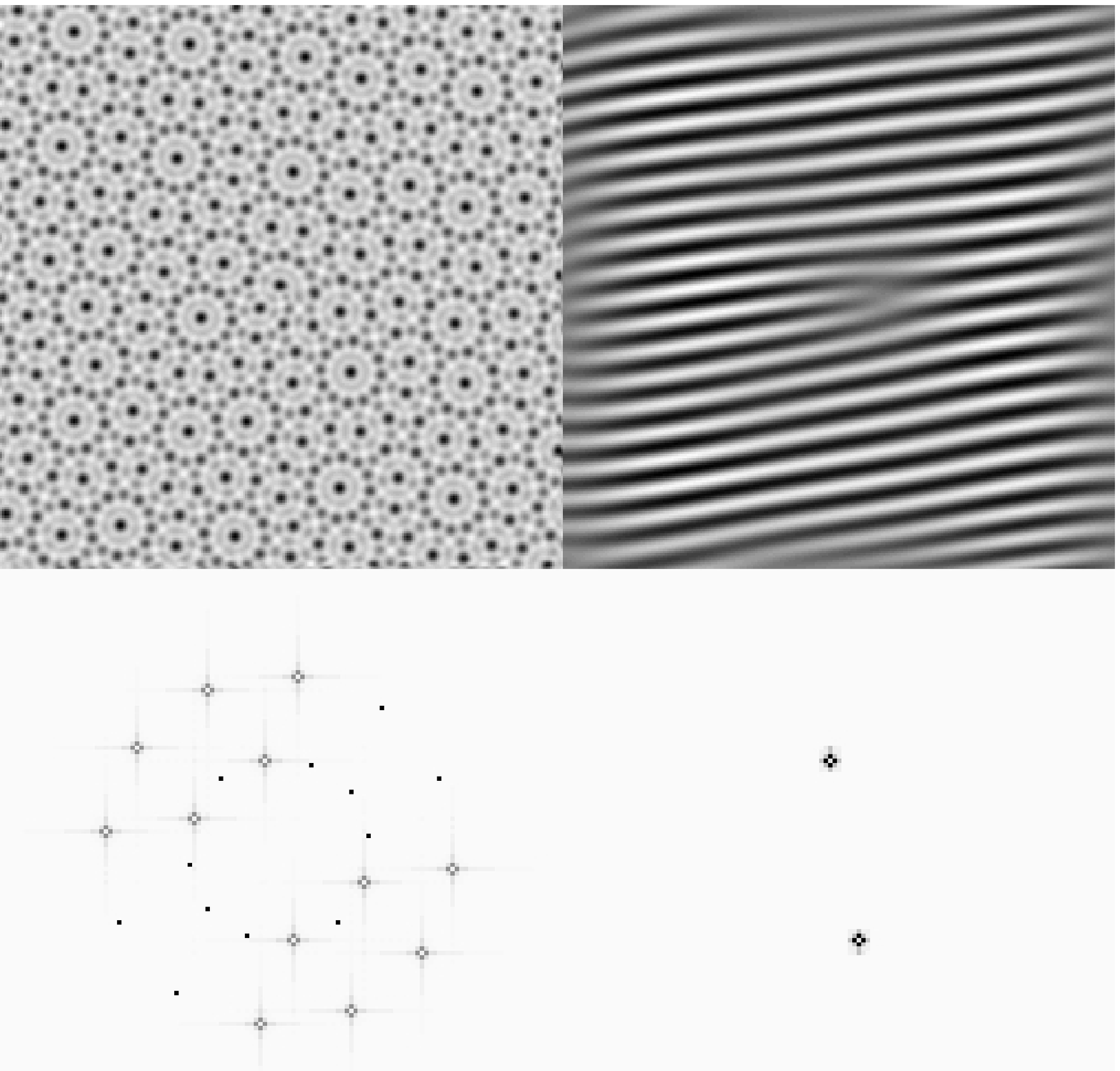}
\end{center}
\caption{Top-left (in both figures): Snapshot of the numerical
      solution of the LP equation(\ref{lpeqn}) showing a square and a
      dodecagonal pattern, a short time after a dislocation has been
      injected. Bottom-left: Fourier transform of the pattern. Note
      the fuzzy Fourier coefficients, containing the information on
      the angle dependent local gauge transformation. Bottom-right: A
      pair of filterred fuzzy Bragg peaks. Top-right: Inverse Fourier
      transform of the filterred peaks clearly showing the dislocation.}
\label{fig:dislocIsolate}
\end{figure}

With this in mind we can easily construct an arbitrary dislocation in
a 2-dimensional periodic or quasiperiodic density. As we traverse in a
loop around the position of the dislocation, say the origin, we
locally change the ground state $\rho(\rv)$ by a gradually varying
gauge function which winds around the $i^{th}$ direction of the
$D$-torus $n_i$ times, and returns back to the original point. This is
most readily accomplished by using an angle-dependent local gauge
function $\chi_\theta(\kv)$ that assigns to the $i^{th}$ basis vector
$\bv^{(i)}$ of $L$ the value $n_i\theta$. Thus, the most general
dislocation is characterized by a set of $D$ integers ($n_1\ldots
n_D$), which for a periodic crystal reduces to the familiar
$d$-dimensional Burgers vector. Not surprisingly, the set of all
dislocations forms a rank-$D$ $\Zset$-module---the so-called homotopy
group of the $D$-torus~\cite{mermindefects}. Examples of
dislocations---injected in this manner into the square and dodecagonal
ground states of the LP equation~(\ref{lpeqn})---are shown in
Fig.~\ref{fig:dislocIsolate} after a short relaxation time.

Once injected into the structure, the positions of dislocations are
tracked numerically, as demonstrated in Fig.~\ref{fig:dislocIsolate},
by filtering individual pairs of Bragg peaks in the Fourier transform,
and then performing an inverse Fourier transform to visualize the
dislocations present in each individual density wave. This allows us
to follow the positions of the dislocations in real time and obtain
quantitative measurements of their velocities as described below.

\begin{figure}[p]
  \begin{center}
    \includegraphics[height=5.2cm]{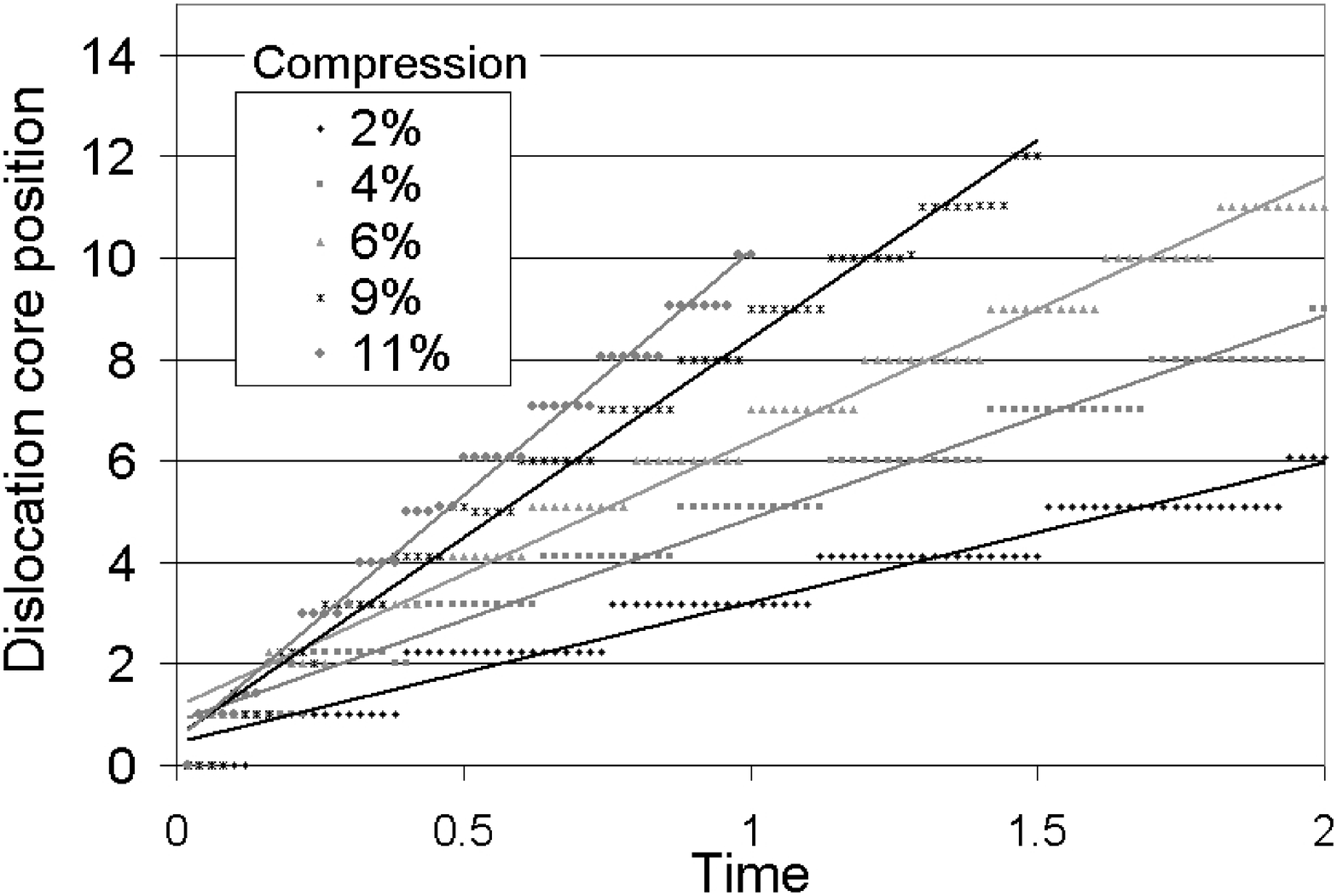}
    \includegraphics[height=5.2cm]{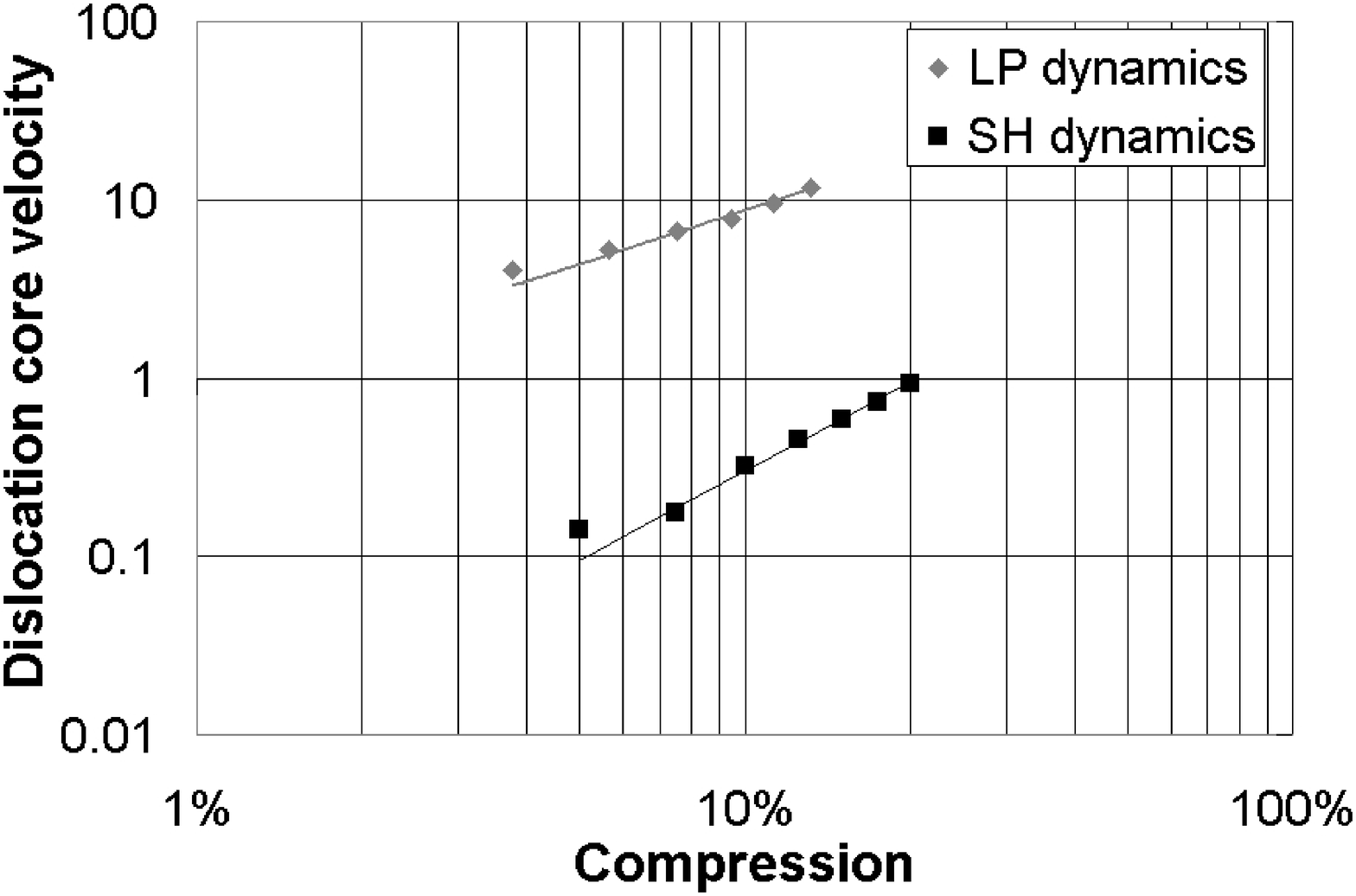}
    \includegraphics[height=5.2cm]{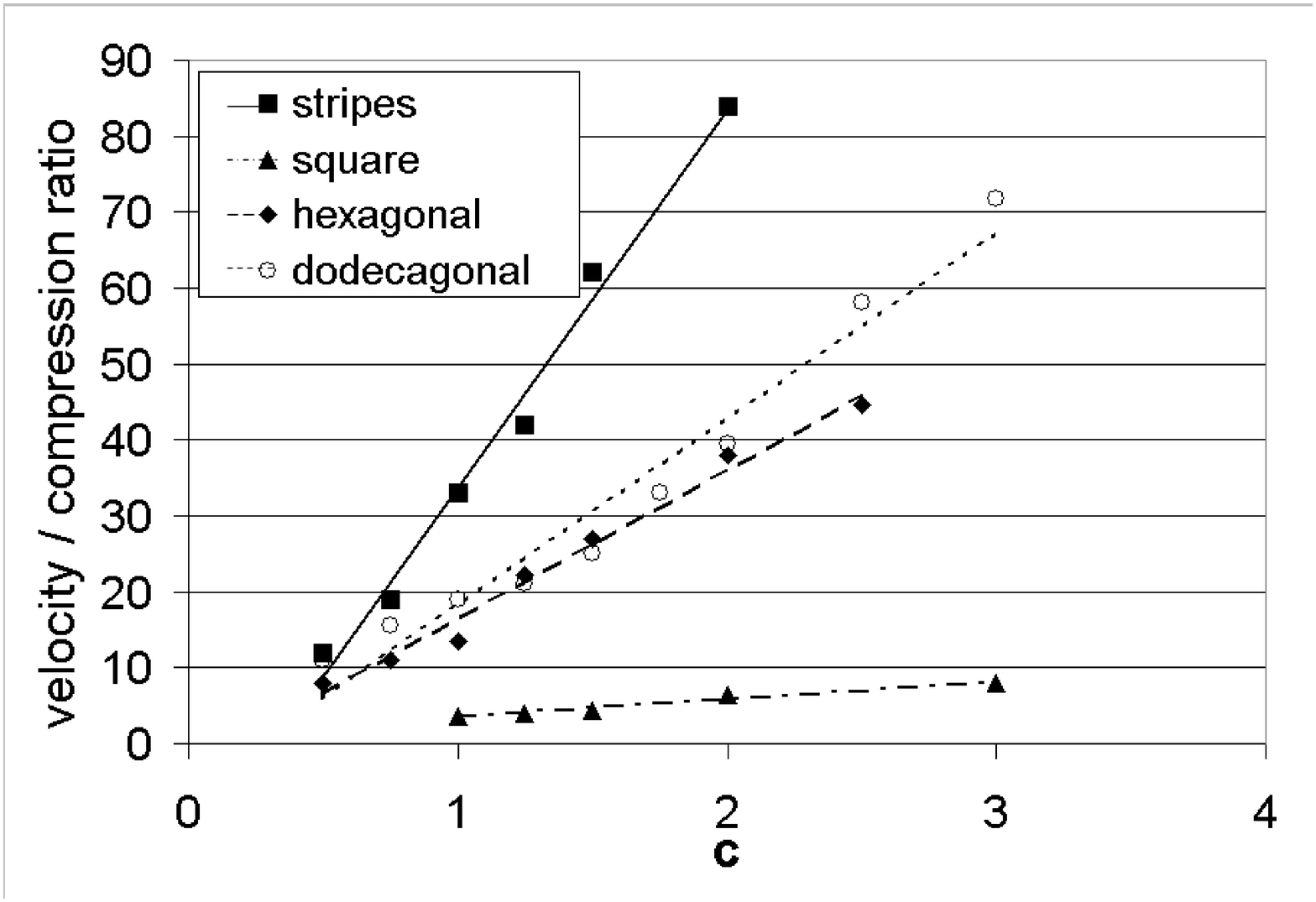}
  \end{center}
  \caption{Measurement of dislocation climb velocity under stress for
    large values of the diffusion constant $c$. (a) Position of a
    dislocation core in a dodecagonal pattern as a function of time
    under Lifshitz-Petrich (LP) dynamics; (b) Dependence of the
    dislocation climb velocity on applied stress in a stripe pattern,
    where $v\propto\delta q^{3/2}$ for Swift-Hohenberg dynamics and
    $v\propto\delta q$ for LP dynamics; (c) Dependence of the
    dislocation climb velocity on the diffusion constant $c$ under LP
    dynamics for different patterns. }
\label{fig:results}
\end{figure}

\section{Dislocation dynamics under stress}

We apply external stress on a structure, containing a single
dislocation, by squeezing it in a particular direction and then
allowing it to evolve under the dynamics of the LP equation. We
quantify the amount of stress by the change $\delta q$ in the
wavenumber of the fundamental density wave in the direction of the
applied stress, relative to its wavenumber in the relaxed steady-state
density. We have calculated elsewhere~\cite{giladthesis}, using the
approach of Siggia and Zippelius~\cite{siggia} and of Tesauro and
Cross~\cite{tesauro}, that under such circumstances, with the
parameters used in our simulations, the dislocation should climb with
a velocity which is linear in the stress $\delta q$ and is
proportional to the parameter $c$, playing the role of a generalized
diffusion coefficient. This should be contrasted with the fact that
for dynamics governed by the Swift-Hohenberg equation~\cite{swift} the
climb velocity is proportional to $\delta q^{3/2}$, for the same
simulation parameters.

Fig.~\ref{fig:results} summarizes our measurements of dislocation
velocities, obtained using the automatic numerical procedure for
tracking the dislocation, described above. Fig.~\ref{fig:results}(a)
shows the position of a dislocation as a function of time for
different values of stress $\delta q$, as measured under the dynamics
of the LP equation for a dodecagonal pattern. For sufficiently large
values of the diffusion constant $c$ the dislocation climbs at a
relatively constant speed, which indeed varies linearly with the
applied stress, as shown in Fig.~\ref{fig:results}(b).

Fig.~\ref{fig:results}(c) compares the measurements for different
steady-state solutions of the LP equation. We find that the stripe
pattern is the easiest for dislocation climb, with the climb velocity
$v \approx 45 c\delta q$, and that the square pattern is the most
resistant, with $v \approx 3 c\delta q$.  Interestingly, the
proportionality constants for the hexagonal and the dodecagonal
patterns are very similar, with $v \approx 25 c\delta q$. A possible
explanation might be that the two triplets of wave vectors making up
the dodecagonal pattern, act independently as two hexagonal patterns
during the climb process.

\section{Dislocation dynamics under stress in the limit of weak diffusion}

\begin{figure}[th]
    \begin{center}
        \includegraphics[width=10cm]{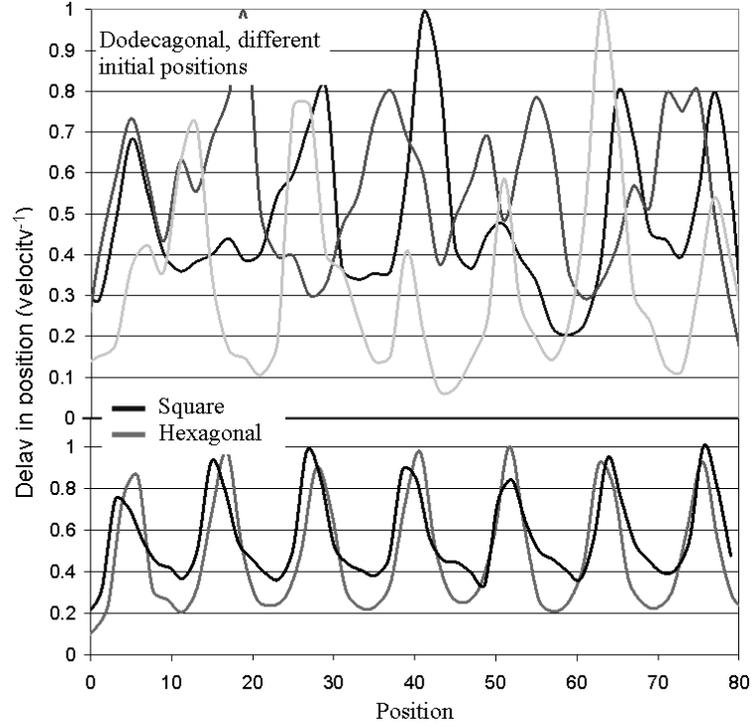}
    \end{center}
    \caption{Bottom: Periodic pinning of a dislocation in a square and
      a hexagonal pattern for a small value of the diffusion constant
      $c$. Top: Irregular pinning of a dilocation in a dodecagonal
      pettern due to the lack of periodicity.}
    \label{fig:pinning}
\end{figure}

As the value of the diffusion constant $c$ decreases the local
features of the pattern become important and the dislocation no longer
climbs at a constant rate. In a periodic pattern, such as the square
or hexagonal densities, the motion of the dislocation nearly comes to
a stop at regularly spaced positions, or pinning sites, as shown at
the bottom of Fig.~\ref{fig:pinning}. These sites correspond to
positions where peaks in the density must first be annihilated before
the dislocation can continue in its climb. In the quasiperiodic
dodecagonal pattern a qualitatively different behavior occurs in which
the dislocation is pinned at irregular intervals, as shown at the top
of Fig.~\ref{fig:pinning}. It seems likely that because the
quasiperiodic pattern contains different local environments the
dislocation is more strongly pinned at certain sites than others, and
these sites never quite repeat. This phenomenon has been observed in
other models of dislocation motion by Mikulla {\it et
  al.}~\cite{mikulla} and Fradkin~\cite{fradkin}.

\medskip
\noindent {\bf Acknowledgment}
This research was supported by the Israel Science Foundation under
Grant No.~278/00.


\begin{thebibliography}{99}
  
\bibitem{urban} K.~Urban, M.~Feuerbacher, M.~Wollgarten, M.~Bartsch,
  and U.~MeSserschmidt, in {\it Physical Properties of Quasicrystals,}
  Z.~M.~Stadnik (Ed.), (Springer-Verlag, Berlin, 1999)
  Chapter 11.
  
\bibitem{trebin} H.-R.~Trebin, in {\it Quasicrystals: An Introduction
    to Structure, Physical Properties and Applications,} J.-B.~Suck,
  M.~Schreiber, and P.~Ha\"ussler (Eds.), (Springer-Verlag, Berlin,
  2002) Chapter 12.
  
\bibitem{woll} M.~Wollgarten, V.~Franz, M.~Feuerbacher, and K.~Urban,
  {\it ibid.}  Chapter 13.
  
\bibitem{faraday} R.~Lifshitz and D.~M.~Petrich, {\it Phys.\ Rev.\ 
    Lett.} {\bf 79} (1997) 1261.

\bibitem{swift} J.~B.~Swift and P.~C.~Hohenberg, {\it Phys. Rev.} A
  {\bf 15} (1977) 319.  
  
\bibitem{jorg} J.~Dr\"ager and N.~D.~Mermin, {\it Phys. Rev. Lett.}
  {\bf 76} (1996) 1489.
  
\bibitem{mermindefects} N.~D.~Mermin, {\it Rev. Mod. Phys.} {\bf 51}
  (1979) 591.
  
\bibitem{giladthesis} G. Barak, {\it M.Sc.~Thesis} (Tel Aviv
  University, 2005).

\bibitem{siggia} E.~D.~Siggia and A.~Zippelius, {\it Phys. Rev. A\/}
  {\bf 24} (1981) 1036.

\bibitem{tesauro} G.~Tesauro and M.~C.~Cross, {\it Phys. Rev. A\/}
  {\bf 34} (1986) 1363.

\bibitem{mikulla} R. Mikulla, P.~Gumbsch, and H.-R.~Trebin, {\it
    Phyl. Mag. Lett.} {\bf 78} (1998) 369.

\bibitem{fradkin} M.~A.~Fradkin, {\it Mat. Sci. Eng. A\/} {\bf
    294-296} (2000) 795.

\end{thebibliography}
\end{document}